\documentclass[3p,times,procedia]{elsarticle}
\usepackage{nupha_ecrc}

\volume{00}

\firstpage{1}

\journalname{Nuclear Physics A}

\runauth{}

\jid{nupha}

\jnltitlelogo{Nuclear Physics A}

\usepackage{amssymb}

\usepackage[figuresright]{rotating}

\begin{document}

\begin{frontmatter}



\dochead{XXVIIIth International Conference on Ultrarelativistic Nucleus-Nucleus Collisions\\ (Quark Matter 2019)}

\title{Local Spin Polarization in 200 GeV Au+Au and 2.76 TeV Pb+Pb Collisions}


\author[a]{Hong-Zhong Wu}
\author[b]{Long-Gang Pang}
\author[c]{Xu-Guang Huang}
\author[a]{Qun Wang}
\address[a]{Department of Modern Physics, University of Science and Technology of China}
\address[b]{Key Laboratory of Quark and Lepton Physics (MOE) and Institute of Particle Physics,
Central China Normal University, Wuhan, Hubei 430079, China}
\address[c]{Physics Department and Center for Particle Physics and Field Theory,
Fudan University, Shanghai 200433, China}

\begin{abstract}
We calculated the azimuthal angle dependence of the local spin
polarization of hyperons in 200 GeV Au+Au and 2.76 TeV Pb+Pb collisions
in the framework of the (3+1)D viscous hydrodynamic model CLVisc with
AMPT initial conditions encoding initial orbital angular momenta.
We find that the azimuthal angle dependence of the hyperon
polarization strongly depends on the choice of the spin
chemical potential $\Omega_{\mu\nu}$. With $\Omega_{\mu\nu}$
chosen to be proportional to the temperature vorticity, our simulation
shows coincidental results with the recent measurements at RHIC.
\end{abstract}

\begin{keyword}
Heavy-ion collision, Local spin polarization, Spin chemical potential, Temperature vorticity

\end{keyword}

\end{frontmatter}


\vspace{0.5cm}

\textit{Introduction.} 
Recently the global polarization of $\Lambda$ (including $\overline{\Lambda}$)
hyperons in non-central heavy-ion collisions has been observed \cite{STAR:2017nature}.
This indicates that the huge orbital angular momentum (OAM) of colliding nuclei
is distributed into the hot and dense medium through the spin-orbit coupling \cite{Liang:2004ph,Gao:2007bc}.
The spin-orbit coupling in parton-parton collisions can be converted to
the spin-vorticity coupling through ensemble average over initial momenta in a fluid
with a shear flow velocity \cite{Zhang:2019xya}. Then the vorticity field leads
to the local hadron polarization along the vorticity direction \cite{Becattini:2013fla,Fang:2016vpj}.
Several theoretical approaches have been developed to study the global
and local polarization in heavy ion collisions based on the assumption
that the spin degree of freedom is in local equilibrium in which the thermal vorticity is involved \cite{Becattini:2013fla,Fang:2016vpj,Pang:2016igs,Florkowski:2017dyn,Weickgenannt:2019dks,Gao:2019znl,Hattori:2019ahi,Wang:2019moi}.

The global polarization effect in the OAM direction can be well
understood by the hydrodynamic and transport models \cite{Karpenko:2016jyx,Li:2017slc,Xie:2017upb,Wei:2018zfb}.
However these models are based on the thermal vorticity and the spin equilibrium assumption
and cannot reproduce the data for longitudinal polarization:
actually there is a sign difference between the data and these model calculations \cite{Becattini:2017gcx,Xia:2018tes,STAR:2019prll}. It worth mentioning that the longitudinal polarization can be described by the the chiral kinetic theory \cite{Sun:2018bjl}
which is for massless fermions instead of massive fermions in the realistic situation.

The assumption that the spin is in a global equilibrium is not always justified,
so the thermal vorticity may not be the right quantity for the spin chemical potential.
In this work, we test different types of spin chemical potentials $\Omega_{\mu\nu}$
and calculate the corresponding local polarization of hyperons.
We use the (3+1)D hydrodynamic model, CLVisc \cite{Pang:2012he,Pang:2018zzo},
to perform the numerical calculation. Same as in Ref. \cite{Wu:2019prr},
we explore four different types of vorticities for $\Omega_{\mu\nu}$ and
calculate the corresponding local polarization with AMPT initial conditions
encoding the global OAM.

\textit{Method. } 
We assume the spin polarization at the freeze-out
hypersurface $\Sigma_{\mu}$ can be expressed as:
\begin{equation}
P^{\mu}(p)=-\frac{1}{4m}\epsilon^{\mu\rho\sigma\tau}p_{\tau}\frac{\int d\Sigma_{\lambda}p^{\lambda}\Omega_{\rho\sigma}n_{F}(1-n_{F})}{\int d\Sigma_{\lambda}p^{\lambda}n_{F}}+O(\Omega_{\rho\sigma}^{2}),\label{eq:polarization-rate}
\end{equation}
where $\Omega_{\rho\sigma}$ is the spin chemical potential,
$p$ denotes the four-momentum of the $\Lambda$ hyperon and $n_{F}=1/[\exp(p_{\mu}\beta^{\mu}-\zeta)+1]$
is its Fermi-Dirac distribution. In the calculation we will set $\zeta=0$
due to the fact that the net baryon density is almost zero
in heavy ion collisions at high energies.

The antisymmetric form of $\Omega_{\rho\sigma}$ is assumed to be
constructed from $T$ and $u^{\mu}$ as
\begin{equation}
\Omega_{\mu\nu}=-(1/2)\lambda(T)[\partial_{\mu}(g(T)u_{\nu})-\partial_{\nu}(g(T)u_{\mu})]\equiv\lambda(T)\omega_{\mu\nu}\label{eq:spin-chemical-potential}
\end{equation}
where $\lambda$ and $g$ are scalar functions of $T$ and $\omega_{\mu\nu}$
is the vorticity tensor. In our calculation four types of vorticities
are considered,
\begin{equation}
\omega_{\mu\nu}^{(K)}=-\frac{1}{2}(\partial_{\mu}u_{\nu}-\partial_{\nu}u_{\mu}),\label{eq:kinematic-vorticity}
\end{equation}
\begin{equation}
\omega_{\mu\nu}^{(\mathrm{NR})}=\epsilon_{\nu\mu\rho\eta}u^{\rho}\omega^{\eta},\label{eq:NR-vorticity}
\end{equation}
\begin{equation}
\omega_{\mu\nu}^{(T)}=-\frac{1}{2}[\partial_{\mu}(Tu_{\nu})-\partial_{\nu}(Tu_{\mu})],\label{eq:Temperature-vorticity}
\end{equation}
\begin{equation}
\omega_{\mu\nu}^{(\mathrm{th})}=-\frac{1}{2}[\partial_{\mu}(u_{\nu}/T)-\partial_{\nu}(u_{\mu}/T)].\label{eq:thermal-vorticity}
\end{equation}
Note that $\omega^{\eta}$ in Eq. (\ref{eq:NR-vorticity}) has the form $\omega^{\eta}=\left(1/2\right)\epsilon^{\eta\alpha\beta\gamma}u_{\alpha}(\partial_{\beta}u_{\gamma})$. For the details of these four types of vorticities, see Ref. \cite{Wu:2019prr}.
As the spin chemical potential should be dimensionless, their explicit forms are 
\begin{equation}
\Omega_{\rho\sigma}^{(i)}=\frac{1}{T}\omega_{\rho\sigma}^{(K)},\frac{1}{T^{2}}\omega_{\rho\sigma}^{(T)},\omega_{\rho\sigma}^{(\mathrm{th})},\frac{1}{T}\omega_{\rho\sigma}^{(\mathrm{NR})},\label{eq:dimensionless-form}.
\end{equation}

When taking an average over the rapidity range $Y\in [-\Delta Y/2,\Delta Y/2]$
and the transverse momentum range $p_{T}\in [p_{T}^{\mathrm{min}},p_{T}^{\mathrm{max}}]$,
we can get the azimuthal angle dependence of transverse and longitudinal
polarization of $\Lambda $ hyperons as:
\begin{equation}
\mathcal{\overrightarrow{\mathcal{P}}}_{i}(\phi_{p})=\frac{1}{\Delta p_{T}}\int_{p_{T}^{\mathrm{min}}}^{p_{T}^{\mathrm{max}}}\left[\frac{1}{\Delta Y}\int_{-\Delta Y/2}^{\Delta Y/2}dYP^{i}(p)\right],\label{eq:phi-dependenced-polarization-rate}
\end{equation}
where $i=x,y,z$ and $P^{i}(p)$ is given by Eq. (\ref{eq:polarization-rate})
and $\Delta p_{T}=p_{T}^{\mathrm{max}}-p_{T}^{\mathrm{min}}$ denotes
the range of the transverse momentum.

\textit{Numerical results. }
For the numerical calculation, we use CLVisc, a (3+1)D relativistic hydrodynamic model \cite{Pang:2012he,Pang:2018zzo} with AMPT initial conditions encoding the global OAM.
The longitudinal polarization can be calculated through
$\left\langle \textrm{cos}\theta_{p}^{*}\right\rangle $
\begin{equation}
\left\langle \textrm{cos}\theta_{p}^{*}\right\rangle =\alpha_{H}\left\langle \textrm{cos}^{2}\theta_{p}^{*}\right\rangle \mathcal{P}_{z},\label{eq:longitudinal-cos-theta}
\end{equation}
where $\theta_{p}^{*}$ is the polar angle of the daughter proton in the $\Lambda(\overline{\Lambda})$'s
rest frame, $\alpha_{H}$ is the hyperon decay parameter ($\alpha_{\Lambda}=\alpha_{\overline{\Lambda}}=0.642\pm0.013$
for $\Lambda$ and $\overline{\Lambda}$), $\mathcal{P}_{z}$ is the longitudinal
component of Eq. (\ref{eq:phi-dependenced-polarization-rate}).

\begin{figure}
\centering{}\includegraphics[scale=0.162]{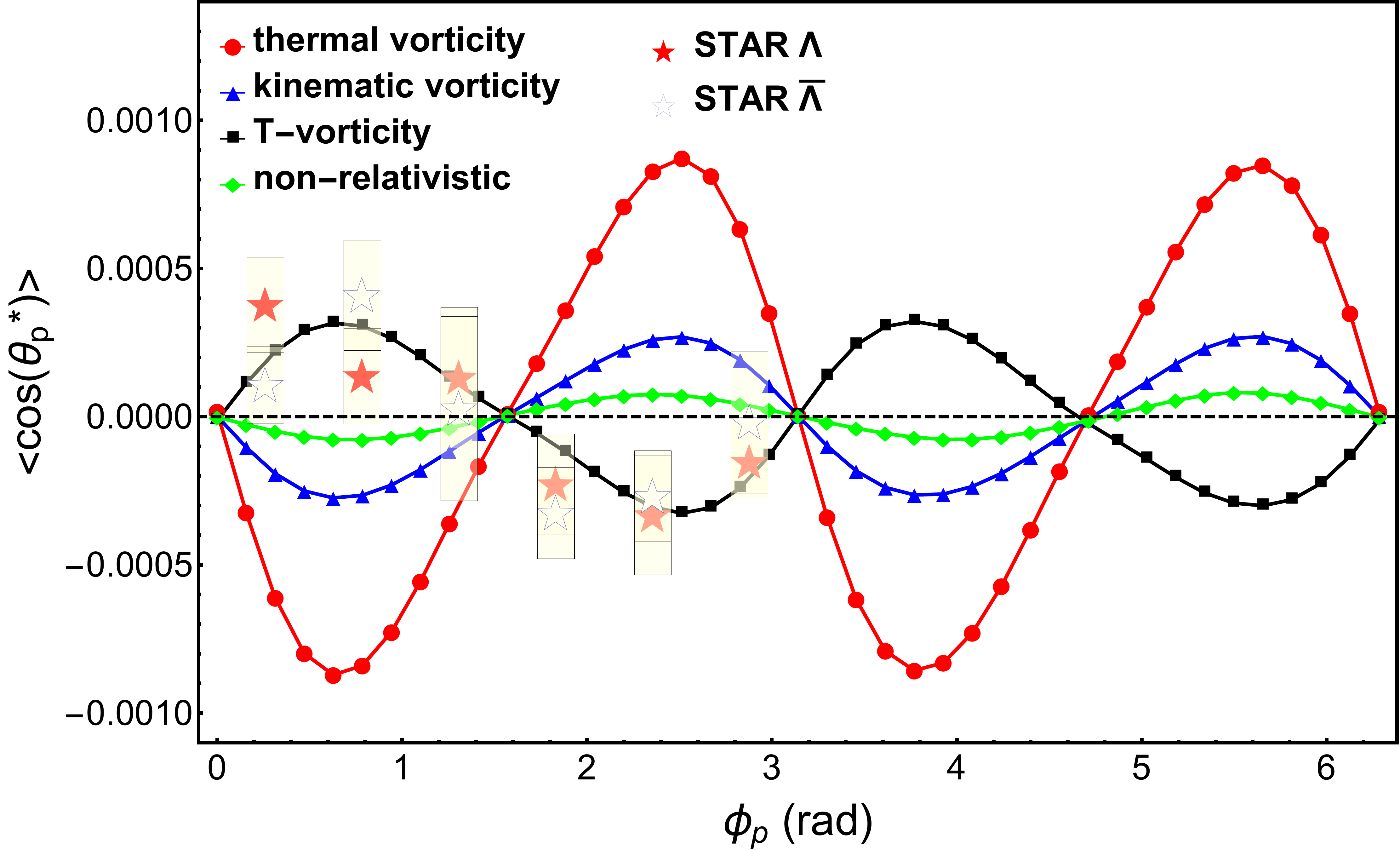}\includegraphics[scale=0.185]{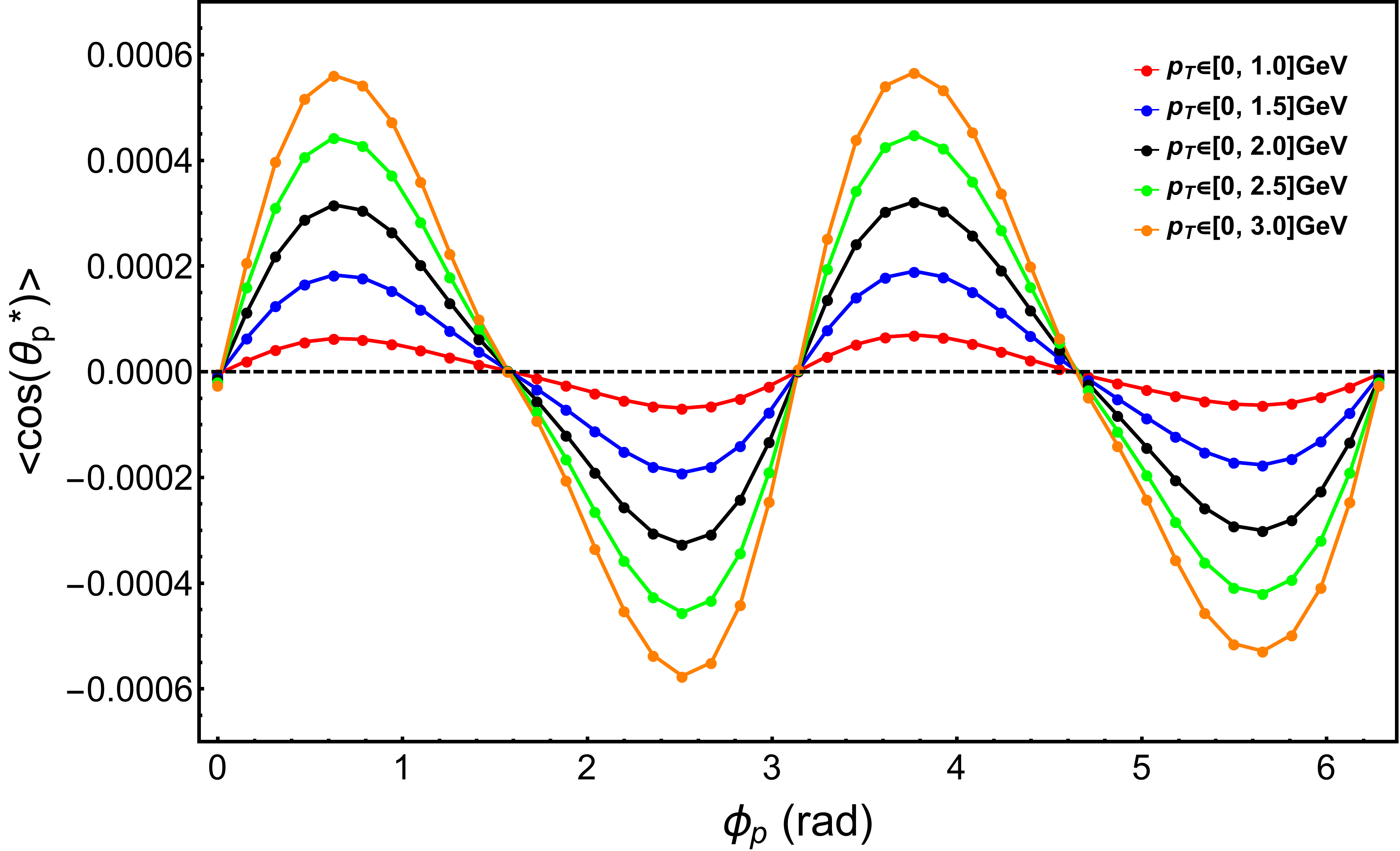}
\caption{\label{fig:AuAu_Pz}The azimuthal angle dependence of longitudinal
polarization in Au+Au collisions at 200 GeV and $20\%-50\%$ centrality.
In our simulation we choose the rapidity range $Y\in [-1,1]$.
The left panel shows the longitudinal polarization for four types of vorticities
and the transverse momentum range $p_{T}\in [0,2.0]$ GeV,
while the right panel shows the dependence of the polarization
with the T-vorticity on different transverse momentum ranges.}
\end{figure}

For Au+Au collisions at $\sqrt{S_{NN}}=200\;\mathrm{GeV}$ and
$20\%-50\%$ centrality, the longitudinal polarization
from four types of vorticities as functions of azimuthal angles
in momentum space are shown in Fig. \ref{fig:AuAu_Pz}.

\begin{figure}
\centering{}\includegraphics[scale=0.185]{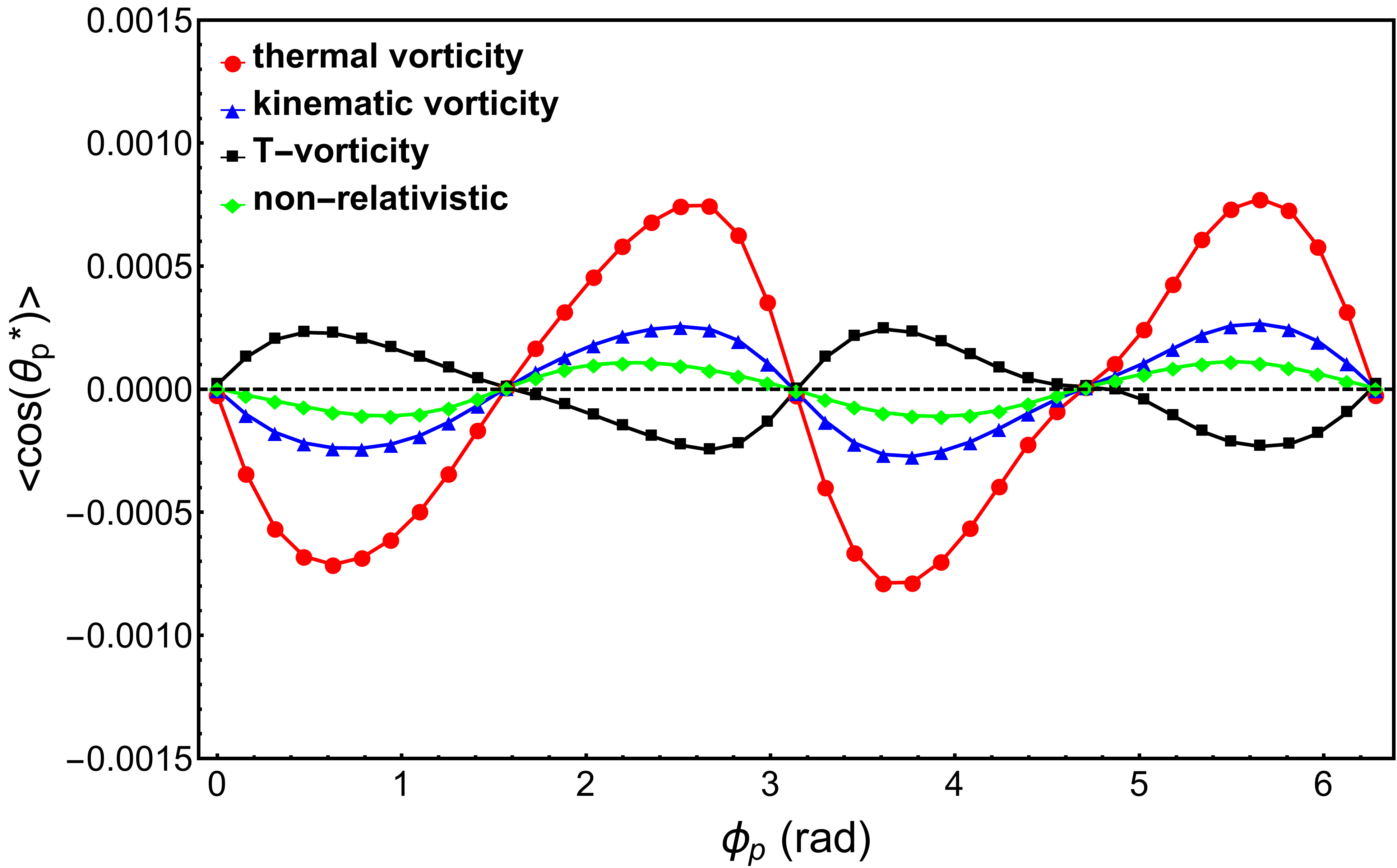}
\includegraphics[scale=0.185]{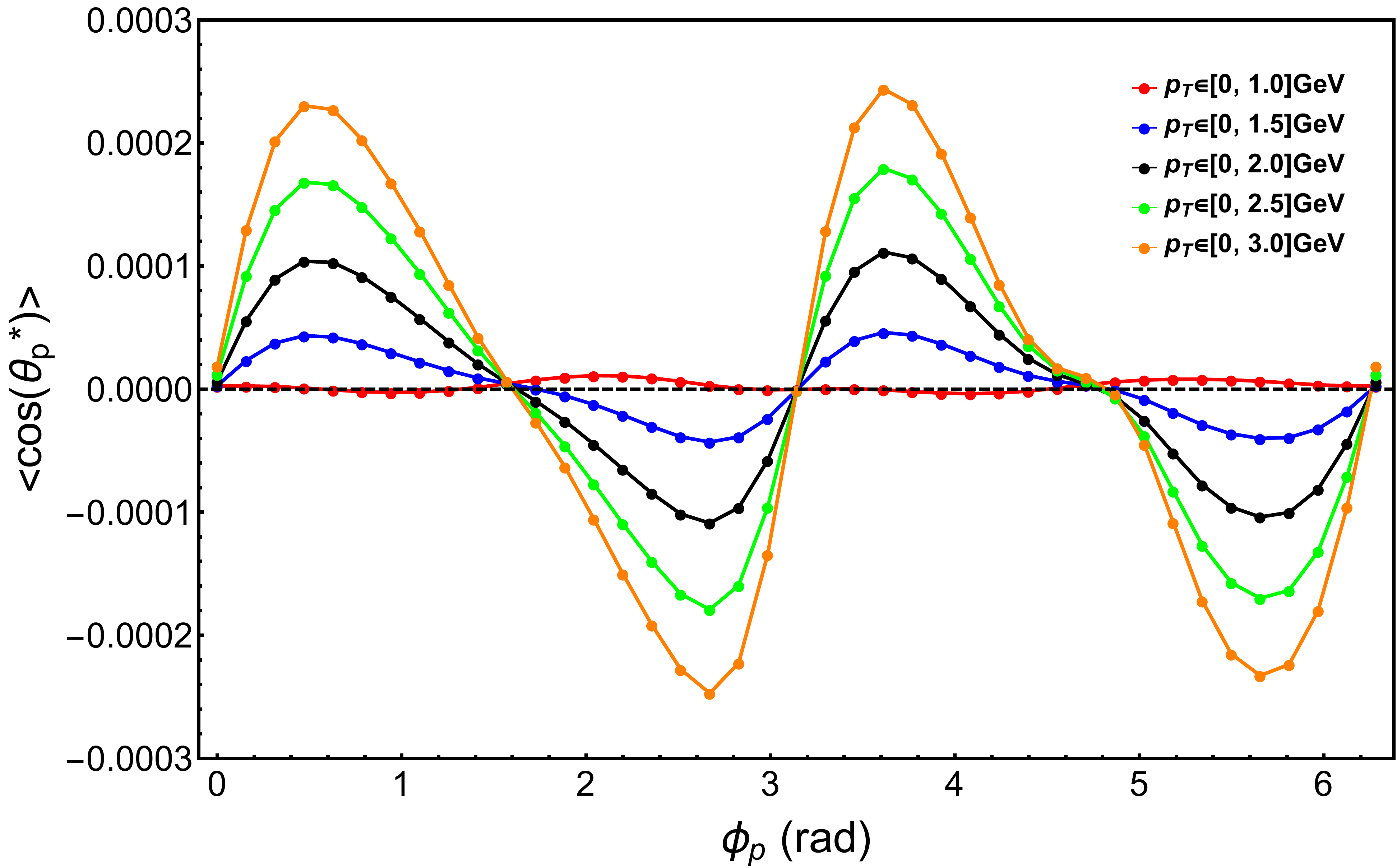}
\caption{\label{fig:PbPb_Pz}The azimuthal angle dependence of longitudinal
polarization in Pb+Pb collisions at 2.76 TeV and $10\%-60\%$ centrality.
In our simulation we choose the rapidity range $Y\in [-1,1]$.
The left panel shows the longitudinal polarization for four types of vorticities and
the transverse momentum range $p_{T}\in [0,3.0]$ GeV, while
the right panel shows the dependence of the polarization
with the T-vorticity on different transverse momentum ranges.}
\end{figure}

\begin{figure}
\centering{}\includegraphics[scale=0.185]{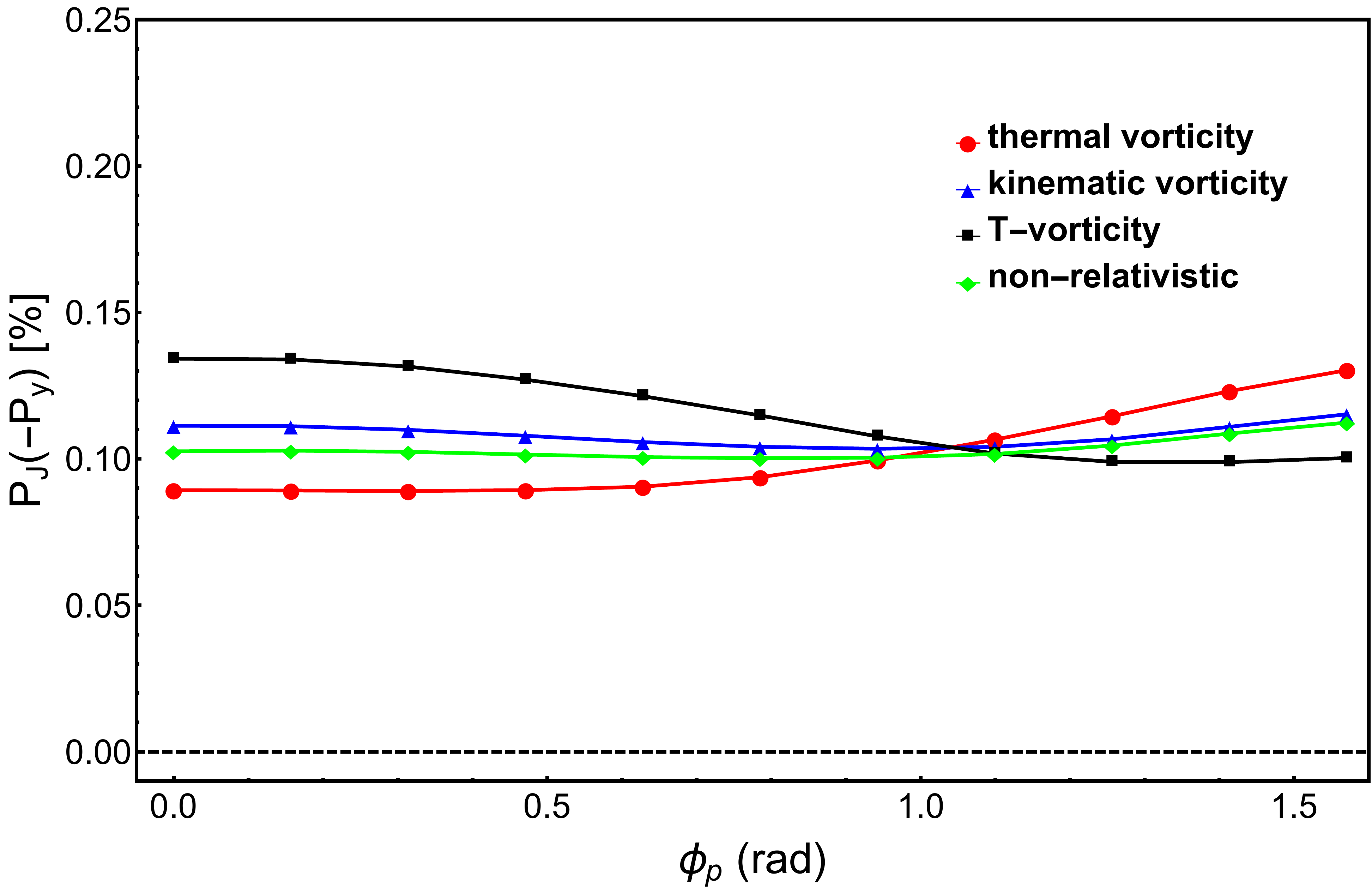}
\caption{\label{fig:PbPb_Py}The azimuthal angle dependence of the polarization
in the direction of the global OAM in Pb+Pb collisions at 2.76 TeV and $10\%-60\%$ centrality.
In our simulation we choose the rapidity range $Y\in [-1,1]$ and $p_{T}\in [0,3]$ GeV.}
\end{figure}

For Pb+Pb collisions at $\sqrt{S_{NN}}=2.76$ TeV and
$10\%-60\%$ centrality, the polarization in the beam direction
and in the OAM direction for four types of vorticities are shown in Fig. \ref{fig:PbPb_Pz}
and \ref{fig:PbPb_Py}, respectively.

In summary, we find that the experimental data of the longitudinal polarization
in Au+Au collisions at 200 GeV can be described quite well by
the T-vorticity. For Pb+Pb collisions at 2.76 TeV, the longitudinal polarization
also has a periodic structure but is smaller than 200 GeV. The magnitude of the
polarization in the direction of the global OAM is consistent with the decreasing
trend in the STAR measurement \cite{STAR:2017nature}.

\textit{Acknowledgement. } 
Q.W. is supported in part by the National Natural Science Foundation of China
(NSFC) under Grants No. 11535012 and No. 11890713.





\bibliographystyle{elsarticle-num}
\bibliography{reference}







\end{document}